# Results from the ULTRA experiment in the framework of the EUSO project

G. Agnetta[a], P. Assis[b], B. Biondo[a,c], P. Brogueira[b], A. Cappa[e], O. Catalano[a,c],
J. Chauvin[d], G. D'Alí Staiti[a,c,f], M. Dattoli[e], M.C. Espirito-Santo[b], L. Fava[e], P. Galeotti[e],
S. Giarrusso[a,c], G. Gugliotta[a,c,†], G. La Rosa[a,c], D. Lebrun[d], M.C. Maccarone[a,c],
A. Mangano[a,c], L. Melo[b], S. Moreggia[d], M. Pimenta[b], F. Russo[a], O. Saavedra[e], P. Scarsi[a,c],
J.C. Silva[b], P. Stassi[d], B. Tomè[b], P. Vallania[g] and C. Vigorito[e] (EUSO Collaboration)
*(a) INAF-IASF, Palermo, Italy*
*(b) LIP, Lisbon, Portugal*
*(c) INFN, Catania, Italy*
*(d) LPSC-Grenoble, France*
*(e) University of Turin and INFN, Torino, Italy*
*(f) DIFTER, University of Palermo, Italy*
*(g) INAF-IFSI and INFN, Torino, Italy*
† *Deceased*
Presenter: P. Vallania (Piero.Vallania@to.infn.it), ita-vallania-P-abs1-he12-poster

The detection of Čerenkov light from EAS in a delayed coincidence with fluorescence light gives a strong signature to discriminate protons and neutrinos in cosmic rays. For this purpose, the ULTRA experiment has been designed with 2 detectors: a small EAS array (ETscope) and an UV optical device including wide field (Belenos) and narrow field (UVscope) Čerenkov light detectors. The array measures the shower size and the arrival direction of the incoming EAS, while the UV devices, pointing both to zenith and nadir, are used to determine the amount of direct and diffused coincident Čerenkov light. This information, provided for different diffusing surfaces, will be used to verify the possibility of detecting from Space the Čerenkov light produced by UHECRs with the EUSO experiment, on board the ISS.

## 1. Introduction

The study of the extreme region of the primary cosmic ray spectrum represents one of the most challenging research in this field. The involved physics items range from astrophysics (acceleration mechanism and propagating space properties) to nuclear and sub-nuclear physics (interaction with the atmosphere nuclei in an energy and pseudo rapidity region very far from actual and future colliders), bearing in mind that we are dealing with particles carrying even 60 Joules in one single particle. At these energies, also very exotic processes, like mini black holes evaporation or high mass particles creation, as Higgs bosons, could be the standard behavior. Due to the very low flux ($< 1$ particle/(km$^2$ century) beyond the Greisen-Zatsepin-Kuzmin cutoff at $E_{GZK} \geq 5 \times 10^{19}$ eV), very large detectors and years of stable operations are required to record a significant amount of data. The Auger project, actually under construction in Argentina, has been designed following this traditional approach, and with 3000 km$^2$ detection area probably represents the ultimate detector of this kind. Despite its gigantic dimensions, only very few events per year are expected at $E > 10^{20}$ eV if the GZK absorption mechanism is present, and also the ultra high energy neutrinos, carrying cosmological informations on the early Universe, cannot be detected with enough statistics. A new detector concept must be exploited, and the observation from Space of the fluorescent and Čerenkov signals produced along the shower development and its impact with the Earth's surface, as proposed by the EUSO Collaboration [1], seems the right answer: with the proposed field of view (f.o.v.) of $\pm 30°$ at the ISS altitude, a surface of $2 \cdot 10^5$ km$^2$ and a target mass of $2 \cdot 10^{11}$ tons of atmosphere can be continuously monitored during night. The feasibility of the fluorescent light detection and the background measurements have been studied elsewhere [2]; the goal of the ULTRA experiment (Uv

Light Transmission and Reflection in the Atmosphere) described in this paper is to verify the possibility of detecting the reflected/diffused Čerenkov light produced by the EAS impacting on different Earth's surfaces (mainly ocean water).

## 2. The experiment

The ULTRA experiment is made by a traditional EAS array, called ETscope, and two different kind of Čerenkov light detectors, called UVscope and Belenos. In order to validate the collected signals, minimizing the possibility of fake events, both direct and diffused Čerenkov light is simultaneously detected, and the diffused signal is seen both from close to the diffusing surface, with a large field detector, and far from the array, to avoid the particle contamination, with a narrow f.o.v. detector. Preliminary measurements have been done at mountain altitude, in Mont-Cenis, to increase the Čerenkov and electromagnetic signals; then we moved to Grenoble (France, 210 m a.s.l.) at the LPSC to perform the detector calibration and optimization. Four ETscope detectors were arranged at the vertex of a square, 54m side, and a fifth detector in the centre to select the events with the shower core located inside the square. The threshold of each detector was set to 0.3 particles; further details on the ETscope detector calibration and performances can be found in [3]. The Čerenkov light detectors were located near the central station: the first pair of Belenos (wide field UV detectors), pointing to zenith, was mounted on ground, while the second one and a couple of Uvscope units (narrow filed UV detectors) were mounted on a scaffold pointing to nadir from a 3.5 m height. The Belenos are equipped with a 20 cm Fresnel borosilicate glass lens, with a second lens 8 cm diameter located in the focus of the first one and used to enlarge the f.o.v. of the device. A 0.5 inch Photomupliplier Tube (PMT) Hamamatsu R960 is located at a distance of 4 cm from the second lens; with this configuration a f.o.v. of $\pm\,40°$ is obtained, with an effective collecting area (taking into account the lens and PMT glass transmission coefficient, the optical configuration and the PMT diameter) increasing from 9 cm$^2$ for para-axial rays to 25 cm$^2$ for the maximum off-axis angle. The PMT is equipped with a BG3 filter to broadly select the lower wavelength region of the Čerenkov spectrum and to reduce the background; the convolution of the filter transmittance, quantum efficiency and glass transmission gives a mean value of 9.4%, with a peak at 330 nm when the Čerenkov spectrum is included. The UVscope optics is made by a single Fresnel acrylic lens, 46 cm diameter, with a 3.5-inches PMT UV-extended in the lens focus. The resulting f.o.v. is $\pm\,4°$.

Due to the high background level and for calibration purposes, a layer of highly diffusing material (Tyvek 1025D, with hemisperical reflectivity $\sim$ 80 % in the 300-400 nm range) is placed in the f.o.v. of the nadir UV detectors. At that time, a double DAQ system was in operation, working simultaneously under LabView DAQ software: a CAMAC system, with a 2229 LeCroy TDC (250 ps/channel) and a 2249W Lecroy ADC charge integrator (300 ns integration time), and a 8 bits, 100 MHz, 6 channels flash ADC, under development at LIP [4]. For this second DAQ system, the total charge and timing are obtained fitting the 10 ns samples; due to the possibility of changing on an event by event basis the integration time to match the effective pulse duration, this system is very effective to evidence very small signals as expected from the UV detectors. A simple calculation of the ratio between the direct Čerenkov signal measured by the Belenos-up and the diffused signal detected by the Belenos-down gives a value of 0.39 for $\theta=20°$ assuming 100% completely diffusing material. This value depends on the shower arrival direction, but it is largely independent from the distance between the detector and the diffusing surface.

## 3. The MC simulation

EAS generated by primary protons have been developed in the atmosphere using the CORSIKA code [5] with the QGSJET hadron interaction model. The showers have been simulated for 5 fixed energies, ranging from $10^{14}$ to $10^{16}$ eV, and arrival direction $\theta = 0°$ and $20°$ (2000 events per point); the observation level has been set to 0 m a.s.l.. In order to reduce the computing time, the NKG option has been used; only a small sub-sample of showers has been simulated with the EGS and Čerenkov options to evaluate the energy thresholds of UV detectors. The CORSIKA output has then been used to calculate the ETscope effective area and shower size resolution: the shower cores have been sampled over a square area increasing with the energy and ranging from 2.9x10$^3$ m$^2$ at 10$^{14}$ eV to 4.7x10$^4$ m$^2$ at 10$^{16}$ eV; the response of each detector has been simulated, including poissonian and experimental fluctuations, and the trigger conditions have been applied; finally the simulated events have been analyzed as the experimental ones fitting the core location and particle density measured by each detector with the usual NKG lateral distribution function. For these studies, only internal events have been chosen, i.e. events with the maximum number of particles detected by the central station; from the experimental data they correspond to 20 % of the total triggers. Fig. 1 shows the convolution of the effective area with the primary spectrum for zenith angles 0° and 20°: the mode of the distribution is $\sim 5 \cdot 10^{14}$ eV, corresponding to the typical detected energy. The detector distances have been indeed optimized to obtain this value, which matches the threshold energy of the Čerenkov light detectors. The shower size and core location resolutions have been calculated using a trial area of 270 × 270 m$^2$ and the reciprocity method [6] to minimize the computing time. At E=10$^{15}$ eV we obtain $\Delta N_e/N_e \sim 35\%$ and $\Delta x_{core}=\Delta y_{core} \sim 6$ m.

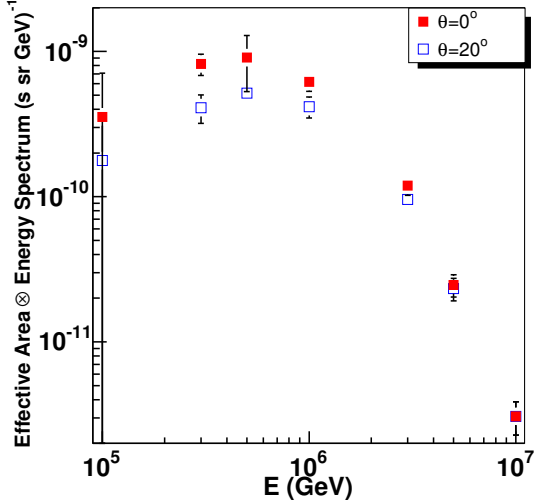
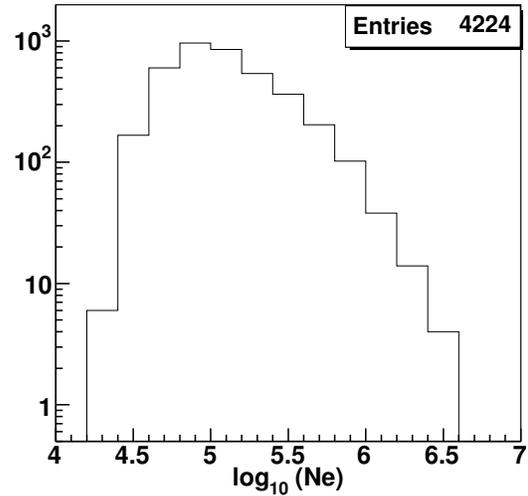

**Figure 1.** Effective area for internal events convoluted with the primary spectrum.

**Figure 2.** Shower size spectrum for internal events.

## 4. Results

Experimental data have been collected from January to December 2004; a total of 21539 events during 948 h of measurements have been recorded. Fig. 2 shows the shower size spectrum for the internal events. The Belenos PMTs have been calibrated using the single photon spectrum; with this calibration and the geometrical acceptance of the optics, the Čerenkov photon density of the incoming shower can be measured with the arrival direction provided by the ETscope array. Since two devices are present for each detector, their coincidence is required; the trigger is given by the 4-fold coincidence of the external ETscope detectors, and the UV devices are acquired passively. Despite the high background level ($\sim$ 3000 ph /(m$^2$ ns sr) in the BG3 filter window, $\sim$ 3-4 times the clear night zenithal background) the required 100 ns coincidence with the central station allows us not to be overwhelmed by random signals. Figure 3 shows the measured photon density at 20 m from the shower core as a function of the shower size N$_e$. Both core location and shower size are determined by the ETscope array; the measured photon density has been scaled to the quoted distance fitting the simulated Čerenkov l.d.f. . Corrections for atmospheric transmission are not applied at this level. The dotted line is the expectation from Monte Carlo simulation for proton induced vertical shower. Even if the fluctuations are big, the agreement between the expected values and the measured mean values is quite good. The square points refer to the CAMAC DAQ, while the circle ones to the LIP DAQ; a slight systematic difference seems to be present, probably due to the different way to process data.

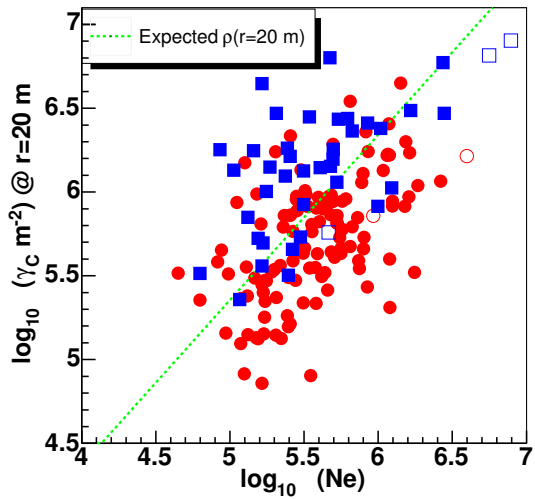

**Figure 3.** Photon density at 20 m from the shower core as a function of the shower size (see text).

The threshold energy can be easily obtained, and corresponds to $\mathrm{Log}_{10}N_e \simeq 4.8$ ($E \simeq 4\times10^{14}$ eV for protons); under this value the direct Čerenkov photons are overwhelmed by the background light and cannot be seen. For diffused light, due to the calculated ratio between up and down, the threshold energy is $\mathrm{Log}_{10}N_e \simeq 5.3$, and, due to the slope of the primary spectrum, we expect that only 14% of the up events have also a diffused signal. The open symbols are the events with both direct and diffused Čerenkov signal: they are 5 over a total of 161, but with different observation times (174.3 h for up and 76.6 h for down). From these events we obtain a mean reflectivity of $77 \pm 17$ %, to be compared with the nominal value of 80 %. Using the down/up background ratio (which is expected to be 1 for 100 % reflectivity) a value of $87 \pm 16$ % is also obtained. We have performed the analysis of the UVscope signals but, due to the large photocathode size of its PMTs and the related position in the array, we cannot be confident that the signals seen by the UVscope detector are entirely due to the diffused Čerenkov light, being the contamination of the shower particles not negligible (particle crossing the PMTs). Such a problem is not present in the Belenos detectors since their photocathode area is about 50 times smaller than the UVscope one.

## 5. Conclusions

The two components of the ULTRA detector (ETscope and Belenos) have been calibrated and tested with real EAS signals, electromagnetic and Čerenkov. Due to the very high background level, the energy threshold for UV light detection is very high, with a very limited number of events. Nevertheless we latched onto the technique. The experiment has been moved to sea surface in Sicily (Capo Granitola, Italy) in April 2005; the measurements have been resumed in May 2005. The ETscope setup is very close to the Grenoble one, and the UVscope is now observing the center of the array far from it, with a very low contamination from shower particles, and the possibility of their discrimination using the timing informations. The first Čerenkov signals diffused by the sea water have already been observed.